\documentclass[12pt]{article}

\usepackage{diagrams}
\def\ind{{\mathchoice {\rm 1\mskip-4mu l} {\rm 1\mskip-4mu l}
{\rm 1\mskip-4.5mu l} {\rm 1\mskip-5mu l}}}

\usepackage{amsmath,amssymb}

\begin{document}
\title{Large-scale games in large-scale systems:\\ tutorial notes}
%
\date{March 27, 2011}
\author{
H. Tembine\footnote{Ecole Superieure d'Electricite, Supelec, France}\thanks{These notes have been prepared for the tutorial course on mean field stochastic games, Supelec, March 2011.}
}

\maketitle

\begin{abstract}  Many real-world problems
modeled by stochastic games have huge state and/or action
spaces, leading to the well-known curse of dimensionality.
The complexity of  the analysis  of large-scale systems is
  dramatically reduced by exploiting  mean field limit and dynamical system viewpoints.   Under regularity assumptions and specific time-scaling techniques, the evolution of the mean field limit can be expressed in terms of deterministic or stochastic equation or inclusion (difference or differential). In this paper,
  we overview recent advances of large-scale games in large-scale systems. We focus in particular on population games, stochastic population games and mean field stochastic games. Considering long-term payoffs, we characterize the mean field  systems using Bellman and Kolmogorov forward equations.
\end{abstract}


\tableofcontents
\section{Introduction}
Dynamic Game Theory deals with sequential situations of several decision
makers (often called  players) where the
objective for each one of the players may be a function of not only
its own preference and decision but also of decisions of other players.

Dynamic games  allow to model sequential decision
making, time-varying interaction, uncertainty and randomness of interaction by the players. They allow to model
situations in which the parameters defining the games vary in
time and the players can adapt their strategies (or policies) according the evolution of the environment.  At
any given time, each player takes a decision
(also called an action) according to a strategy. A (behavioral) strategy of a player is a collection of history-dependent maps that tell at each time the choice (which can be probabilistic) of that player.
 The
vector of actions chosen by players at a given time  may determine not only the payoff for each player at
that time; it can also determine the state evolution.
 A particular class of dynamic games widely studied in the literature is the class of {\it stochastic games}. Those are dynamic games with probabilistic state transitions (stochastic state evolution) controlled by one or more players. The discrete time state evolution is often modeled as interactive Markov decision processes while the continuous time state evolution is referred to stochastic differential games. Discounted stochastic games have been introduced in \cite{shapley53}.
Stochastic games and interactive Markov decision processes  are widely used for modeling sequential
decision-making problems that arise in engineering, computer science,
operations research,  social sciences etc. However, it is well
known that many real-world problems modeled by stochastic games  have huge state
and/or action spaces, leading to the well-known {\it curse of dimensionality} that
makes solution of the resulting models intractable. In addition, if the size of the system grows without bound, the number of parameters: states, actions, transitions explode exponentially.

In this paper we present recent advances in large-scale games in large-scale systems. Different models (discrete time, continuous, hybrid etc) and different coupling structures (weakly or strongly) are presented.
Mean field solutions are obtained by identifying
a consistency relationship between the individual-state-mass interaction such that in the population
limit each individual optimally responds to the mass effect and these individual strategies also
collectively produce the same mass effect presumed initially. In the finite horizon case, this leads to a coupled system forward/backward optimality equations (partial differential equation or difference equations).

\subsection{Structure}
The remainder of the paper is structured as follows. In the next section we overview the mean field model description and its wide range of applications in large-scale wireless networks. We then focus on different mean field coupling formulation. After that we present mean field related approaches. The novelties of the mean field systems are discussed.
\subsection{Notations}
We summarize some of the notation used in the paper in Table \ref{tablenot3}.
\begin{table}[htb]
\caption{Summary of Notations} \label{tablenot3}
\begin{center}
\begin{tabular}{ll}
\hline
  Symbol & Meaning \\ \hline
  $f_t$ & drift function (finite dimensional) \\
  $\sigma_t$ & diffusion function at time $t$\\
  ${x}^n_{j,t}$ &  state of player $j$ in a population of size $n$\\
  $q_{xux',t}$ & transition probability at time $t$\\
  $M^n_t$ & mean field process \\
  $\mathcal{L}_{xx',t}(u,m)$ & transition kernel of the population profile\\
  ${x}_{j,t}$ & limit of state process ${x}^n_{j,t}$ \\
  $r_t$ & instantaneous payoff function\\
  $g_T$ & terminal payoff function
  \\ \hline
\end{tabular}
\end{center}
\end{table}

\section{Overview of large-scale games}
\subsection*{ Population games }
 Interactions  with large number of   players and different types can be described as a sequence of dynamic games.
Since the population profile involves many  players for each type or class and location, a common approach is to replace
individual players and to use continuous variables to represent the aggregate average of {\it type-location-actions}. The validity of this method has been proven only under specific time-scaling techniques and regularity assumptions. The {\it mean field limit} is then modeled by state and location-dependent time process. This type of aggregate models are also known as non-atomic or population games. It is closely related to von Neumann (1944) and mass-action interpretation in Nash (1951). In the context of transportation networks, interactions between continuum of players  have been studied by Wardrop (1952) in a deterministic and stationary setting of identical players.

In finite game, a (Nash) equilibrium is characterized by $\forall j,$
  $$
  \overline{\{ x_{j}\in\mathcal{X}_j,\ m_{j,x_j}>0\}}=\mbox{support}(m_j)\subseteq\arg\max_{x'_j\in\mathcal{X}_j} r_j(e_{x'_j},m_{-j})
  $$ where $r_j(.)$ denotes payoff of $j,$ $\mathcal{X}_j$ its action space and $m_j$ its randomized action, $m_{-j}=(m_{j'})_{j'\neq j}.$

  In the infinite population game, a (Nash) equilibrium is characterized by a fixed inclusion: the support of the population profile is included in argmax of the payoff function,
$$
\overline{\{ x\in\mathcal{X},\ m_{x}>0\}}=\mbox{support}(m)\subseteq\arg\max_{x'\in\mathcal{X}} r_{x'}(m).
$$
In other words, if the fraction of players under a specific action is non-zero then the payoff of the corresponding action is maximized. This
large-scale methodology has inherent connections with evolutionary game theory when one
is studying a large number of interacting players in different subpopulations. Different solution concepts such as evolutionarily state states or strategies, neutrally stable strategies, invadable states have been proposed and several applications can be found in evolutionary biology, ecology, control design, networking and economics.

\subsection*{ Overview of mean field stochastic games }
We briefly present related works on mean field stochastic games.

$\bullet$
Discrete time mean field stochastic games with continuum of players have been studied by \cite{jovanovic} under the name  {\it anonymous sequential games}. The authors considered the evolution of the mean field limit in the Bellman dynamic programming equation.
 The work in  \cite{jovanovic} shows, under suitable conditions,  the existence of such mean field equilibria
in the case where the  mean field limit of players' characteristics evolves nonstochastically.
 The authors in \cite{bergin} showed how {\it stochastic mean field limit}
can be introduced into the model (so the mean field limit evolves stochastically).

$\bullet$
 Decentralized stochastic mean field control and {\it Nash Certainty Equivalence} have been studied in \cite{m1,m2,caines1,caines2,meyn} for large population stochastic dynamic systems. Inspired by mean field approximations in statistical mechanics and linear quadratic Gaussian (LQG) differential games, the authors analyzed a common situation where the dynamics and payoffs (costs,reward, utility) of any given agent are influenced by certain aggregate of the mass multi-agent behaviors and established the existence of optimal response to mean field under boundedness and regularity assumptions. In the infinite population limit, the players become statistically independent under some technical assumptions on the control laws and the structure of state dynamics, a phenomenon related to the {\it propagation of chaos} in mathematical physics. In \cite{MRP}, the authors   extended the LQG mean field model to non-linear state dynamics and non-quadratic case  for localized and multi-class of players.  LQG hybrid mean field games have been considered  in \cite{zhang2}.

$\bullet$   In \cite{lasry1,lasry2,lasry3}    a  mathematical modeling approach for
highly dimensional systems of evolution equations corresponding to a large number of players (particles or
agents) have been developed. The authors extended the field of such mean-field approaches also to problems in economics, finance and game theory. They studied $n$-player stochastic differential
games  and the related problem of the existence of  equilibrium points, and by letting $n$
tend to infinity they derived the mean-field limit equations such as Fokker-Planck-Kolmogorov (FPK) equation coupled with the mean field version of Hamilton-Jacobi-Bellman-Fleming (HJBF). Applications to finance can be found in \cite{geant}. The authors in \cite{buck2,buch,zhang} extended the framework  to mean field stochastic differential games under general structure of drift and noise function but also with major and minor players.
 The authors in \cite{christian2010,aime} applied mean field games to crowd and pedestrian dynamics. Numerical methods for solving backward-forward partial differential  equations  can be found in~\cite{bf2010}.

 $\bullet$ Discrete time models with many firm dynamics  have been studied
by  \cite{weintraub,weintraub2}  using decentralized strategies. They  proposed
the notion of oblivious equilibria via a mean field approximation. Extension to unbounded cost function can be found in \cite{johari}.
In \cite{ramesh10}, a  mean field equilibrium analysis of dynamic games with complementarity structures have been conducted.
In \cite{tembinephd,tem1}, models of interacting players in discrete time with finite
number of states have been considered. The players share local resources. The
players are observable only through their own state which changes according to a Markov decision process. In the limit, when the number of players goes
to infinity, it is found that the asymptotic system is given by a non-linear dynamical system (mean field limit). The mean field limit can be in discrete or in continuous
time, depending on how the model scales with the number of players. If the expected number
of transitions per player per time slot vanishes when the size of the system grows, then the limit is in continuous
time. Else the limit is in discrete time. Markov mean field teams have been studied \cite{tembinephd}, Markov mean field  optimization, controls and Markov decision processes have been studied in \cite{meanfieldlectures}.
Connection of the resulting limiting mean field games to {\it anonymous games} or
stochastic population games have been established. A stochastic population game given by a population profile
which evolves in time, internal states for each player and a set of actions in each state and
population profile. The expected payoff of the player  are completely determined by the population
profile and its current internal state. At the continuum limit of the population, one can have (i)  a discrete time mean field games which cover the so-called {\it anonymous sequential games} or (ii) a continuous time mean field games leading the so-called {\it differential population games}. The corresponding limiting games fall down to

(i)  Differential population games in which the optimality criteria leads an extended
HJBF coupled with FPK equations or,

(ii)
  {\it Anonymous sequential  games} in which the leading dynamics are mean field version of Bellman-Shapley equations combined with discrete time mean field Kolmogorov forward equations similar to the  prescribed dynamics developed by  \cite{jovanovic}.

\subsection*{Networking applications}

Below we present the relevance of large-scale games in large-scale networks.
 Due to the limitations of the classical perfect simulation approaches in presence of large number of entities, mean field approach can be more appropriate in some scenarios:

  \subsubsection*{ MFSG and continuum modeling} The  simulation of multiple networks and their statistical modelling can be very expensive, whereas solving a continuum equation such as partial differential equation can be less expensive in comparison. Example of such large-scale systems include:
\begin{itemize}\item Internet of things with 2 billions of nodes,
\item Network of sensors deployed along a volcano, collecting large quantities of data to monitor seismic activities where transmissions are from relay-node to relay-node until finally delivered to a base station
\item Disruption-tolerant networks with opportunistic meeting in a large population of 20.000.000 nodes
    \end{itemize}
%



\subsubsection*{ Opportunistic interaction under random mobility: }
 The work in \cite{christian2010,aime} has modelled crowd behavior and pedestrian dynamics using a mean field approach.
 Inspired from \cite{christian2010}, one can get a random mobility model for the users.
In \cite{meanfieldlectures} an application
to malware propagation in opportunistic networking have been studied. This example illustrates how mean field game dynamics can be useful
in describing the network dynamics in absence of  infrastructure, low connectivity and in absence of fixed routes to disseminate information. The model has been extended to Brownian mobility of players with communication distance parameter and energy saving in wireless ad hoc networks.
A challenging problem of interest such in configuration is a routing packet over the wireless  network from  sources to  destinations (their locations are unknown and they can move randomly). The wireless random path maximizing the quality of service with minimal end-to-end delay from a source to a destination changes continuously as the network traffic and the topology change. An expected element characterizing the network state (mean field) and mean field learning-based routing protocol are therefore needed to estimate  the network traffic and to predict the best network behavior.

\subsubsection*{ MFSG for  carrier sense multiple protocols: }
The mean field stochastic game approach has potential applications in wireless networks  (see \cite{johari} and the references therein).
Mean field Markov models have been studied in details in \cite{mama2,gossip} for Carrier Sense Multiple Access (CSMA)-based IEEE 802.11 Medium Access Control (MAC) protocols and gossip protocols. When the strategies of the users are taken into consideration, one gets interdependent decision processes for the backoff stage:
The backoff process in IEEE 802.11 is governed by a Markovian decision process
if the duration of per-stage backoff is taken into account:
\begin{itemize}\item every node in backoff state $x_{\theta}$ attempts transmission with
probability $\frac{1}{\gamma n+\beta_2+\beta_3\ln(n)}u_{x_{\theta}}^{\theta}$ for every time-slot; \item if it succeeds, the backoff state changes
to 0; \item otherwise, the backoff state changes to $(x_{\theta} + 1)$ mod $(K_{\theta} + 1)$
where $K_{\theta}$ is the index of the maximum backoff state in class $\theta.$
\end{itemize}
Extension to SINR-based admission control and quality of service (QoS) management with backoff state can be found \cite{meanfieldlectures}.

\subsubsection*{ Mean field power allocation }
In \cite{meanfieldlectures} the authors study a power management problem using mean field stochastic game. The mean field approach have been applied to dynamic power allocation (vector) in green cognitive radio networks. The authors showed that if the players react to the mean field and, if  the size of the system is sufficiently large then decentralized mean field power allocations can be approximated equilibria.

MFSG for energy market in smart grid, chemical reaction and water composition and molecular mobility can be found in \cite{meanfieldlectures}

\section{Basics of MFSG models} \label{t3sec2}
In this section we overview basics  of mean field stochastic game (MFSG) models.
\subsection{Weakly coupling}
{\bf Weakly coupling via the payoff functions}
The players are weakly coupled only via the payoff functions if the individual state dynamics are not directly influenced by the others states and strategies i.e
\begin{equation}x^n_{j,t+1}=\bar{f}_{j,t}^n(x^n_{j,t},u^n_{j,t},w^n_{j,t})
\end{equation} where
$x^n_{j,t}$ is the state of player $j,$ $\bar{f}^n_{j,t}$ is a deterministic function, $u^n_{j,t}$ is the action/control of player
$j$ and $w^n_{j,t}$ is a random variable (independent to the state and the action processes of others) with transition probabilities given by $$\mathbb{P}(x_{t+1}\in \bar{X}| x^n_{t},u^n_{j,t},\ldots,u^n_{j,0},x^n_{j,0}),$$ where $\bar{X}$ is a subset of $\mathcal{X}.$
The instantaneous payoff function of player $j$ may depend on the state and/or actions of the others or the state mean field $\frac{1}{n}\sum_{j=1}^n \ind_{\{x_{j,t}^n=x \}}  $ or the state-action mean field $$\frac{1}{n}\sum_{j=1}^n \ind_{\{(x_{j,t}^n,u^n_{j,t})=(x,u) \}}  $$ or the population profile process
$
\frac{1}{n}\sum_{j=1}^n \delta_{x_{j,t}^n},
$ etc.

Note that in dynamic environment, the players may not interact all the time with the same set of neighbors. Some players may be active or inactive, some new player may join or leave the game temporary etc. Then the payoff function depends on the state and also the actions of all the players that she/he meets during the long-run interaction.

A simple  continuous time version of the above state dynamics is the following It\^o stochastic differential equation (SDE)
\begin{equation}
dx^n_{j,t}=f^n_{j,t}(x^n_{j,t},u^n_{j,t}) dt +\sigma^n_{j,t}(x^n_{j,t},u^n_{j,t} ) d\mathcal{B}_{j,t}
\end{equation}
where $\sigma^n_{j,t}$ is the variance function and $f^n_{j,t}$ is the drift function for player $j$ at time $t$ and $\mathcal{B}_{j}$ is a standard Brown motion (Wiener process). An example of such dynamics is $dx^n_{j,t}=u^n_{j,t}dt+\sigma^n_{j,t}  d\mathcal{B}_{j,t}$

{\it How the payoff depends on the mean field?}
When the number of players is very large, the payoff function can be expressed in function of the mean field under technical conditions. Here is a simple example. Let the instantaneous payoff  functions be in the following form
$$
r^n_{j,t}=\frac{1}{n}\sum_{i=1}^n \bar{r}^n_{j,t}(x^n_{j,t},u^n_{j,t},x^n_{i,t}).
$$
Let recall that for any measurable bounded function $\phi$ defined over the state space, one has
\begin{eqnarray}
\int_w \phi(w)\left[ \frac{1}{n}\sum_{i=1}^n\delta_{x^n_{i,t}}\right](dw)
=\int_w \phi(w)M^n_t(dw) =
\frac{1}{n}\sum_{i=1}^n \phi(x^n_{i,t})
\end{eqnarray}
Thus, the instantaneous payoff function is
$$
r^n_{j,t}(x^n_{j,t},u^n_{j,t},M^n_t)=\int_w\bar{r}^n_{j,t}(x^n_{j,t},u^n_{j,t},w)M^n_t(dw).
$$

The long-term payoff function can be in finite horizon or in infinite horizon with discount factor or not.

{\bf Weakly coupling via the individual states}
 Here we focus on the case where the players are  only weakly coupled via the individual states. In this case, the payoff functions of each player depends only its own state and own strategy but also his/her state is influenced by the other players states and actions.

An example of such discrete time  dynamics is
\begin{equation}x^n_{j,t+1}=\bar{f}_{j,t}^n(x^n_{j,t},u^n_{j,t},x^n_{-j,t},u^n_{-j,t},w^n_{j,t})
\end{equation}
where transition kernel of $w^n_{j,t}$ depends on the states and the actions of the others: $\mathbb{P}(.| x^n_{t},u^n_{t},\ldots,u^n_{0},x^n_{0})$ where $x^n_{-j,t}=(x^n_{j',t})_{j'\neq j}, x^n_t=(x^n_{j,t})_{j},  u^n_t=(u^n_{j,t})_{j},\ t\geq 0.$

An example of continuous time version is
\begin{equation}
dx^n_{j,t}=f^n_{j,t}(x^n_{j,t},u^n_{j,t},x^n_{-j,t},u^n_{-j,t}) dt +\sigma^n_{j,t}(x^n_{t},u^n_{t} ) d\mathcal{B}_{j,t}
\end{equation}
which covers the following dynamics:
\begin{eqnarray}
dx^n_{j,t}=\frac{1}{n} \sum_{i=1}^n \bar{f}^n_{j,t}(x^n_{j,t},u^n_{j,t},x^n_{i,t},u^n_{i,t}) dt\nonumber \\ +\frac{1}{n} \sum_{i=1}^n\bar{\sigma}^n_{j,t}(x^n_{j,t},u^n_{j,t},x^n_{i,t},u^n_{i,t} ) d\mathcal{B}_{j,t}
\end{eqnarray}

The case where $f^n_{j,t}$ and $\sigma^n_{j,t}$ depend only the state are well-studied. Then, the averaging structure becomes
\begin{equation}
dx^n_{j,t}=\frac{1}{n} \sum_{i=1}^n \bar{f}^n_{j,t}(x^n_{j,t},u^n_{j,t},x^n_{i,t}) dt +\frac{1}{n} \sum_{i=1}^n\bar{\sigma}^n_{j,t}(x^n_{j,t},u^n_{j,t},x^n_{i,t}) d\mathcal{B}_{j,t}
\end{equation}

The last equation can be written in function of the mean field $M^n_t=\frac{1}{n}\sum_{j'=1}^n \delta_{x^n_{j',t}}:$
\begin{eqnarray}
dx^n_{j,t}=\left[\int_w \bar{f}^n_{j,t}(x^n_{j,t},u^n_{j,t},w)M^n_t(dw)\right] dt \nonumber \\  +\left[ \int_w \bar{\sigma}^n_{j,t}(x^n_{j,t},u^n_{j,t},w )M^n_t(dw)\right] d\mathcal{B}_{j,t}
\end{eqnarray}
For discrete time models, the similarity  with the above methodology can be done in the transition probabilities.
 Another way is to consider directly the model in which the probabilities depend  on the fraction of players with specific state by considering $\frac{1}{n}\sum_{j=1}^n \ind_{ \{x^n_{j,t}=x\}}.$ If  the transitions depend only on a local mean field then it can written as a function of mean field seen from that player.

{\bf Weakly coupling via neighborhoods}
Consider the individual dynamics in the form:
$$
\left\{
\begin{array}{c}
dx_{j,t}^n=\sum_{i\in\mathcal{N}_{j}}\omega^n_{ij}(t) f^n_{\theta_j,t}(x_{j,t}^n,u^n_{j,t},x_{i,t}^n,u^n_{i,t}) dt\\ + \sum_{i\in\mathcal{N}_{j}}\omega^n_{ij}(t) \sigma^n_{\theta_j,t}(x_{j,t}^n,u^n_{j,t},x_{i,t}^n,u^n_{i,t}) d\mathcal{B}_{j,t},\\
x_{j,0}^n\in \mathcal{X}\subseteq\mathbb{R}^{k},\ k\geq 1\\
j\in\{1,2,\ldots,n\}, \theta_j\in\Theta\\
\end{array}
\right.
$$ where coefficient $\omega^n_{ij}(t)\geq $ represents the influence of player $i$ to player $j$ at time $t$. Then, player $j$ has its own local mean field limit $M^n_{j,t}:=\sum_{i\in\mathcal{N}_{j}} \omega^n_{ij}\delta_{(x^n_{i,t},u^n_{i,t})}$ where
$n$ is the number of players, $x^n_{j,t}$ is the state of player $j$, $u^n_{j,t}$ is the control of player $j,$ $\mathcal{B}_{j}$ is a standard Brown motion (Wiener process),
the coefficients are normalized such that
$$
\omega^n_{ij}\geq  0,\ \sum_{i\in\mathcal{N}_j} \omega_{ij}^n=1.
$$
Then $\omega^n_{ij}=0$ can be interpreted as the case where player $i$ does not affect the state dynamics of player $j.$ The term $\theta_j$ is the type of the player $j.$ $\Theta$ is the set of types.

Then, under suitable conditions, the asymptotic of a subsequence of the individual state dynamics lead to macroscopic McKean-Vlasov equation with local mean field limit under the form:
$$
\left\{
\begin{array}{c}
dx_{j,t}=\int_{w'}f_{\theta_j,t}(x_{j,t},u_{j,t},w')\ m_{j,t}(dw') dt\\ + \int_{w'}\sigma_{\theta_j,t}(x_{j,t},u_{j,t},w')\ m_{j,t}(dw') d\mathcal{B}_{j,t},\\
x_{j,0}^n\in \mathcal{X}\subseteq\mathbb{R}^{k},\ k\geq 1\\
u_{j,t}\in\mathcal{U}_{\theta_j}
\end{array}
\right.
$$ Note that the processes $m_{j,t}$ are interdependent and their laws can be obtained as a solution of coupled systems of Fokker-Planck-Kolmogorov equations.
Moreover, the convergence rate is in order of $O(\frac{1}{\sqrt{n}}+\epsilon^0_n)$ where $\epsilon^0_n$ captures the initial estimates and the gap at the initial distributions.
We refer the reader to \cite{meanfieldlectures} for more recent discussions on the convergence issue.

\subsection{Strongly coupling}
Here the state evolutions and the payoff functions depend on the state and/or the strategies of some of the other players. Typically, most of games with variable number of interacting during time fall down in the class of strongly coupling mean field interaction. For example, the instantaneous payoff
$$r^n_{j,t}=\sum_{i\in\mathcal{N}_{j}}\omega^n_{ij}(t) \bar{r}^n_{\theta_j,t}(x_{j,t}^n,u^n_{j,t},x_{i,t}^n,u^n_{i,t})
$$ and the state dynamics
$$\left\{
\begin{array}{c}
dx_{j,t}^n=\sum_{i\in\mathcal{N}_{j}}\omega^n_{ij}(t) f^n_{\theta_j,t}(x_{j,t}^n,u^n_{j,t},x_{i,t}^n,u^n_{i,t}) dt\\ + \sum_{i\in\mathcal{N}_{j}}\omega^n_{ij}(t) \sigma^n_{\theta_j,t}(x_{j,t}^n,u^n_{j,t},x_{i,t}^n,u^n_{i,t}) d\mathcal{B}_{j,t},\\
x_{j,0}^n\in \mathcal{X}\subseteq\mathbb{R}^{k},\ k\geq 1\\
j\in\{1,2,\ldots,n\}, \theta_j\in\Theta\\
\end{array}
\right.
$$ lead to a strongly coupling mean field interaction.
\section{What is new?}
The novelties of the MFSG approach are in the characterization of the mean field optimality\footnote{Note that ``mean field optimality'' refers to response to a consistent mean field. It is not necessarily optimal in the finite regime.}. Theses optimality equations differ from the classical dynamic games and dynamic programming principles.
\subsection{Discrete time}
In the mean field stochastic Markov game modeling in discrete time, there must  be an equation to express
the dynamic optimization problem of each player. Usually this involves one equation for
each player. If players are classified together by similar player types, there is
one equation per type. This equation is generally a Bellman-Shapley equation,
since a large proportion of dynamic optimization problems with perfect state observation fall within the framework of
dynamic programming. Hence,  the Bellman-Shapley
equations will be used to compute optimal behavioral strategies.
An equation is also needed to express the subpopulations' behavior, the mean field behavior of each type.  The
dynamics of the distribution is governed by a Kolmogorov forward  equation. In the Kolmogorov forward equation, the optimal behaviors of the players occur as data, since it is the infinite collection of
individual behaviors that is aggregated and  constitutes
collective behavior by consistency.
Thus, the modeling of  the behavior of a group of players naturally leads to
a BS-K (Bellman-Shapley  and Kolmogorov)  system of equations. The discrete BS-K  have been studied by Jovanovic \& Rosenthal in the eighty's. The novelty in their study is that the
mean field
games formalism involves {\it the density of players on the state space
can enter in the Bellman-Shapley equation}. Thus, the mean field equilibrium is defined by an BS-K system
in which the Bellman-Shapley equations are doubly coupled: individual
behaviors are given for the Kolmogorov forward equation  and, at the
same time, the distribution of players in the state space enters in the Bellman equation which is completely innovative. This means that players can incorporate into their preferences the density of
states/actions of other players at the anticipated equilibrium. Therefore each player
can construct his strategy by taking account of the anticipated distribution of
strategies and of the actions of other players. Under suitable conditions, this fixed-point of behaviors, the mean field equilibria can be defined by moving to the
limit on the number of players in the class of Markov games in discrete time (or  difference games) that are asymptotically invariant by permutation within the same type of players called {\it Asymptotic Indistinguishability Per Class}\footnote{These assumptions follow the line of the works by de Finetti (1931), Hewitt \& Savage (1955), Aldous (1983), Sznitman (1991), Graham (2000), Tanabe (2006), McDonald (2007) etc.}.

\subsection{Continuous time}
In the continuous time model, the Hamilton-Jacobi-Bellman-Fleming (HJBF) equation will replace the Bellman equation and the Kolmogorov forward equation becomes a Fokker-Planck-Kolmogorov (FPK) equation. We then get a coupled system of partial differential equations (PDEs). In addition, in presence specific player such major player, its individual state dynamics at the limit regime should be added to the system. Then, the question of existence, uniqueness, regularity, and performance bounds arise for the system of PDEs. See the mean field games (MFG) lectures by Lions at College de France.
\subsection{Connection between the mean field models}
The reader may ask what is the connection between all the above  mean field models.

{\it Is there a connection between the discrete time Markov model and the mean field differential game model?}

The authors in \cite{meanfieldlectures} give a partial answer to this question. Under particular structure of payoff functions and probability transitions of the mean field stochastic population game model one can get a {\it mean field differential game} at the limit for vanishing intensity of interactions. This establishes a first connection from discrete time to continuous time mean field model. Next, we need to show that the convergence of subsequences of optimal strategies and optimal payoffs under the Bellman-Shapley's equation to the Hamilton-Jacobi-Bellman equation under mean field dynamics.
 The authors provided sufficient conditions for mean field stochastic games with random number of interacting players for mean field convergence to stochastic differential equations. Their techniques for the mean field optimality criterion  combine  It\^o-Dynkin's formula with stochastic maximum principle.

 A second connection can be obtained by considering {\it mean field stochastic difference game}. Under specific time-scales, one show that the discrete time mean field stochastic game converges to a mean field stochastic differential game characterized by a non-linear macroscopic McKean-Vlasov, Fokker-Planck-Kolmogorov and HJBF equations.

Following the same setting, one can design numerical scheme of the It\^o stochastic differential to move from differential mean field model into {\it difference mean field model}. But still one needs to show  that the strategies, the values, $\epsilon-$Nash properties holds under these scaling schemes because these properties depends mainly on the proposed scheme for the time-derivative and integration of the partial differential equations (PDE).

\section{Mean field related approaches} \label{t3sec3}
In this section we present mean field related approaches.
\subsection{Connection to mathematical physics}
There are connections between exact microscopic models that govern the evolution of large
particle systems and a certain type of approximate models known in Statistical Mechanics as
mean field limit. This notion of mean field limit is best understood by getting acquainted
with the most famous examples of such equations inspired from physics. The particle system model describes
the evolution of a generic player (particle) subject to the collective interaction created by a
large number $n$ of other players (particles). The state of the generic player is then given by
its phase space density; the force field exerted by the $n$ other players on this generic player is
approximated by the average with respect to the phase space density of the force field exerted
on that particle from each point in the phase space. A number of models have been studied
in the literature. Those are McKean-Vlasov equation, Fokker-Planck equations, mean-field
Schr\"{o}dinger equation, Hartree-Fock equation, Bergers equation, Boltzmann equations, transport equations, continuity equations etc.

Incorporation of controls in these models give controlled mean field equations. If in addition a dynamic optimization setting were present, one gets a large-scale dynamic  game.

\subsection{Connection to evolutionary  dynamics}
The paradigm of evolutionary game dynamics  has been to associate relative growth rate to
actions according to the expected payoff they achieved,
then study the asymptotic trajectories of the state of the system,
i.e. the fraction of players that adopt the different individual and actions.
The works in \cite{kurtz70,book,benaim03,Bill}  derive {\it mean field game dynamics} for multiple-type population games.
These mean field game dynamics are  generalization of  {\it evolutionary game dynamics} (deterministic or stochastic). For large populations with finite number of states and/or actions in $\mathcal{X},$ the standard deterministic evolutionary game dynamics  based on revision protocols  are in the form
\begin{equation}
\dot{m}_t(x)=\sum_{x'\in\mathcal{X}}\mathcal{L}_{xx'}(m_t)m_{t}(x')-m_t(x) \sum_{x'\in\mathcal{X}}\mathcal{L}_{x'x}(m_t)
\end{equation} which is a specific Kolmogorov forward equation.  The term $\mathcal{L}_{xx'}$ represents a rate transition from $x$ to $x'.$

 This equation can be obtained from the drift limit  and single selection per time unit without control parameter (\cite{Bill,tembine09,isdg2010}). By specifying the transitions probabilities $\mathcal{L}$, one gets
  Replicator dynamics, Best-response dynamics, Smith dynamics, Brown-von Neumann-Nash dynamics,
  Orthogonal projection dynamics, Target projection dynamics, Ray-projection dynamics, Smooth best response dynamics, Imitative Boltzmann-Gibbs dynamics, Multiplicative weight imitative dynamics, Generalized pairwise comparison dynamics, Excess payoff dynamics, ``Imitate the better'' dynamics  etc. See \cite{manzoorjsac,manzoorinfocom,manzoormagazine}

\subsection{Connection to the propagation of chaos}
  If the mean field stochastic games model satisfies the invariance in law by any permutation with players index within the same type under specific controls $u$ that preserve this property, one can use the {\it  exchangeability per class} or indistinguishability per class \cite{graham08} to establish a propagation of chaos \cite{Sznitman,Tanabe}. Let $x^n_j=(x^n_{j,t})_{t\geq 0}.$ Then, the process $\Lambda^n=\frac{1}{n}\sum_{j=1}^n\delta_{x^n_j}$ converges in law to a random process $\tilde{m}$ with law $\mu$  is equivalent to the so-called $\mu-$chaoticity: for any integer $k,$ any measurable and bounded functions $\phi_1,\ldots,\phi_k$
\begin{equation}
\lim_n\ \mathbb{E}\left(\prod_{j=1}^k \phi_j(x^n_j)  \right)=\prod_{j=1}^k \left(  \int_{w\in\mathcal{X}}\phi_j(w) \mu (dw)\right)
\end{equation}
{\it Non-commutative diagram}
Consider a population with $n$ players. Denote the mean field by $M^n_t=\sum_{j=1}^n\omega^n_j\delta_{x^n_{j,t}}$ where $x^n_{j,t}$ is the state of player $j$ at time $t$ and $\omega^n_j$ is the weight of player $j$ in the hull population of size $n.$ Then, given a initial condition $m_0,$ denote by  $M^n_t[u,m_0]$ the process $M^n_t$ starting with the distribution given by $m_0$ at time $0$ subject to the control $u.$ The study of the process $M^n_t[u,m_0]$ is summarized in the following diagram:

\begin{diagram}
M^{n}_t[u,m_0] & \rTo^{t\longrightarrow +\infty} & \varpi^n[u,m_0] \\
\dTo_{n\longrightarrow +\infty} & & \dTo_{n\longrightarrow +\infty} \\
m_t[u,m_0] & \rTo^{t\longrightarrow +\infty} & ?\\
\end{diagram}

If the limits are well-defined, we call $\varpi^n=\lim_t\ M^n_t$ and $m_t=\lim_n M^n_t.$ Then, the question is
on the double limit i.e the commutativity  of the diagram.

It turns out that the double limit can be different. The diagram is not always commutative.
$$
\lim_n \lim_t M^n_t \neq \lim_t \lim_n M^n_t.
$$

This phenomenon is in part due to the fact that the stationary distribution of the process $\varpi^n$
 is unique under irreducibility conditions and the dynamics of $m_t$ may lead to a {\it limit cycle}. As
a consequence, many techniques and approaches based on stationary regime (such as fixed point
techniques, limiting of frequencies state-actions approaches in sequence of stochastic
games, replica methods, interacting-particle systems etc) need some justification. This difference in the double limits  (the non-commutativity
phenomenon) suggests to be careful about the use of stationary population state
equilibria as the outcome prediction and the analysis of equilibrium payoffs since this equilibrium
may not be played. Limit cycles are sometimes more appropriate than the stationary
equilibrium approach.

The convergence to an independent and identically distributed system is sometimes referred to {\it  chaoticity}, and
the fact that chaoticity at the initial time implies chaoticity at further times is called propagation of chaos.
This diagram says that, in general the chaoticity property may not holds in {\it stationary regime}. This means that two randomly picked players in the population may be correlated.

 We mention a particular case where the rest point $m^*$ is related to the $\delta_{m^*}-$ chaoticity.
If the mean field dynamics of $m_t$ has a unique global attractor $m^*$ then the propagation  of chaos property holds for the measure $\delta_{m^*}.$ Beyond this particular case, one can have multiple rest points but also the double limit $\lim_n \lim_t M^n_t$  may differ from  $\lim_t \lim_n M^n_t$  leading a non-commutative diagram. Thus, a deep study of the dynamical system is required if one want to analyze a performance metric for a stationary regime. A {\it counterexample}   of different double limits is provided in \cite{meanfieldlectures}.

\subsection{Weak convergence}
$\bullet$ {\it de Finetti-Hewitt-Savage}
Consider a complete separable metric space $\mathcal{X}$ and a sequence of random processes $(x^n_{j})_{j,n},$ satisfying the indistinguishability per class property i.e invariance in law of permutation within the same type/class. Then, the population profile $M^n$ converges weakly to a random measure $m.$ Moreover, conditionally to $m,$ one has
that for any integer $k,$ any measurable and bounded functions $\phi_1,\ldots,\phi_k$ defined over $\mathcal{X},$
\begin{equation}
\lim_n\ \mathbb{E}\left(\prod_{j=1}^k \phi_j(x^n_{j})  \ | \ m \right)=\prod_{j=1}^k \left(  \int_{w\in\mathcal{X}}\phi_j(w) m (dw)\right)
\end{equation}

$\bullet$ Now we focus on the convergence of the pair $(x^n_{j,t},M^n_t).$ In the case where $M^n_t$ goes to a deterministic object $m_t,$ vanishing time-scales, it is shown in \cite{tem1} that the pair $(x^n_{j,t},M^n_t)$ converges weakly to $(x_{j,t},m_t)$ where $x_{j,t}$ is a continuous time jump and drift process (which depends on $m$) $m_t$ is a solution of an ordinary differential equation.
\subsection{Differential population game}
In this subsection we provide a mean field equilibrium characterization of the {\it differential population game}~\cite{tembinephd} where each generic player reacts to the mean field for a finite horizon $[0,T].$
We first start by a payoff of the form $\bar{r}_t(u,m).$
$$(*)\ \sup_u\ [\bar{g}_T(m_T)+\int_t^T \bar{r}_{t'}(u_{t'},m_{t'})\ dt']$$ subject to the mean field dynamics $${m}_t=m_0+\int_0^t \tilde{f}_{t'}(u_{t'},m_{t'})\ dt'.$$

We say the pair of trajectories $(u^*_t,m^*_t)_{t\geq 0}$ constitutes a consistent mean field response if $u^*_t$ is an optimal strategy to be above  problem (*) where $m^*_t$ is the mean field at time $t$ and $u^*_t$ produces the mean field  $m_t[u^*,m_0]=m^*_t$

A consistent mean field response  is characterized by a backward-forward equation
$$
\left\{
\begin{array}{c}
\bar{v}_T(m)=\bar{g}_T(m)\\
-\partial_t \bar{v}_t=\sup_{u}\left\{ \bar{r}_t(u,m_t)+\langle \nabla_m \bar{v}_t, \tilde{f}_t(u,m_t)\rangle \right\}
\\
m_t=m_0+\int_0^t \tilde{f}_{t'}(u^*_{t'},m_{t'})\ dt'
\end{array} \right.
$$ where $u^*_t$ is in argmax of $\bar{r}_t(u,m_t)+\langle \nabla_m \bar{v}_t, \tilde{f}_t(u,m)\rangle.$

Next, we consider a individual state-dependent payoff $r_t(x,u,m).$ Define $$F^{1}_T(x,u,m)=g_T(x_T,m_T)+\int_t^T r_{t'}(x_{t'},u_{t'},m_{t'})\ dt'$$ where $g_T$ is a terminal payoff.
$$ (**) \sup_u\ [g_T(x_T,m_T)+\int_t^T r_{t'}(x_{t'},u_{t'},m_{t'})\ dt']$$
subject to the mean field dynamics $${m}_t=m_0+\int_0^t \tilde{f}_{t'}(u_{t'},m_{t'})\ dt'.$$
where the individual state  $x_t=x_t[u]$ is a continuous time Markov  jump process under $u.$ We denote by $\bar{q}$ the infinitesimal generator of $x_t[u].$ See \cite{tem1,tembinephd} for more details on the analysis of the process $(x_{j,t},m_t).$

We say the pair of trajectories $(u^*_t,m^*_t)_{t\geq 0}$ constitutes a  mean field equilibrium if $\{u^*_t\}_{t\geq 0}$ is a mean field response to be above  problem (**) where $m^*_t$ is the mean field  at time $t$ and $u^*_t$ produces the mean field  $m_t[u^*,m_0]=m^*_t$

 Consider a differential population game problem with  single type. Assume that there exists a unique pair
 $(u^*,m^*)$ such that

(a) there exists a bounded, continuously differentiable function
 $\tilde{v}_x:\ [0,T]\times \mathbb{R}^{|\mathcal{X}|},\ \tilde{v}^*_{x,t}(m)=v_t(x,m)$  and differentiable function $m^*:\ [0,T] \longrightarrow \mathbb{R}^{|\mathcal{X}|},$ $m^*_t=m_t[u^*,m_0]$ solution to the  backward-forward equation:
$$
\left\{
\begin{array}{c}
v_T(x,m)=g_T(x,m),\\
-\partial_t v_t(x,m)=\sup_{u}\left\{ r_t(x,u,m)+\langle \nabla_m v_t(x,m), \tilde{f}_t(u^*,m)\rangle \right. \\ \left.+\sum_{x'\in\mathcal{X}}\bar{q}_{xux',t}(m) v_t(x',m)\right\}
\\
m_t=m_0+\int_0^t \tilde{f}_{t'}(u^*_{t'},m_{t'})\ dt'\\
x_0=x\in\mathcal{X},  m_0\in\ \Delta(\mathcal{X})
\end{array}
\right.
$$

(b) $u^*_t(x)$ maximizes of the function $$r_t(x,u,m_t)+\langle \nabla_m v_t(x,m), \tilde{f}_t(u^*,m_t)\rangle+\sum_{x'\in\mathcal{X}}\bar{q}_{xux',t}(m_t) v_t(x',m)$$   where $\bar{q}_{xux',t}(m)$ is the transition of the infinitesimal generator of $x_t$ under the strategy $u$ and $m,$ $\sum_{x'}\bar{q}_{xux',t}(m)=0,$ the term $\sum_{x'\in\mathcal{X}}\bar{q}_{xux',t}(m) v_t(x',m)$ is $$\sum_{x'\neq x} \bar{q}_{xux',t}(m) (v_t(x',m)-v_t(x,m)),$$ $m_t[u^*,m_0]=m^*_t$

Then,
$(u^*_t,m^*_t)_{t\geq 0}$ with $m^*_t=m_t[u^*,m_0]$ constitutes a mean field equilibrium and $ \tilde{v}^*_{x}(m^*)= v(x,m^*)=F_{x,T}(u^*,m^*).$

Similarly, for multiple types the systems becomes
$$\left\{
\begin{array}{c}
v_{\theta,T}(y_{\theta},m)=g_{\theta,y_{\theta}}(m),\\
\!\!\!-\partial_t v_{\theta,t}(x,m)=\displaystyle{\sup_{u_{\theta}}}\left\{ r_{\theta,t}(y_{\theta},u_{\theta},m){+}\langle \nabla_m v_{\theta,t}(x,m), \tilde{f}_{\theta,t}(u^*,m)\rangle \right. \\ \left.+\sum_{y'_{\theta}}\bar{q}_{y_{\theta}u_{\theta}y_{\theta}'}(m) v_{\theta,t}(y'_{\theta},m)\right\}
\\
m_{\theta,t}=m_{\theta,0}+\int_0^t \tilde{f}_{\theta,t'}(u^*_{t'},m_{t'})\ dt'\\
y_{\theta,t}=y_{\theta}, \ m_0\in\ \Delta(\mathcal{X}),
\theta\in\Theta.
\end{array}
\right.$$

Note that the applicability of this result is limited  because in general the $\arg\max$  may not be reduced to a singleton.

\section{Mean field systems}
\subsection{ Transition kernels:}
In this subsection we briefly present the mean field systems. In the discrete time case, the BS-K equation for finite horizon $T$  is given by
 $$
\left\{
\begin{array}{c}
m_{t+1}(x')=\sum_{x}m_{t}(x) \mathcal{L}_{xx',t}(u^*_t,m_t)\\
\forall t, x,a \ \mbox{such that}\ m_t(x) u^*_t(a|x)>0 \Longrightarrow\\
a\in\arg\max_{b} \left\{r_t(x,b,m_t)+\sum_{x'}v_{t+1}(x',m_{t+1})q_{xbx'}(u_t,m_t)
\right\}
\end{array}
\right.
$$ Under regularity and boundedness of the instantaneous payoff and the transition kernels, the existence of solutions of the backward-forward system can be established using Kakutani-Glicksberg-Fan-Debreu fixed point theorem.

\subsection{ Mean field It\^o's SDE}
In this subsection we present the backward-forward system for mean field limit satisfying It\^o's stochastic differential equation.
The mean field system for horizon $T$ in continuous time for a payoff in the form $\mathbb{E}\left( g_T(m_T)+\int_0^T r_t(u_t,m_t)\ dt\right)$ is given by (McK-V-FPK):
 $$\left\{\begin{array}{c}\tilde{v}_T(m)=g_T(m) \\
-\frac{\partial}{\partial t}\tilde{v}_t=\sup_{u_t\in\mathcal{U}}\left\{r_t(m_t,u_t)\right. \\ \left. +\sum_{x\in \mathcal{X}}\tilde{f}_{t,x}(m_t,u_t)\frac{\partial }{\partial m_x}\tilde{v}_t\right. \\ \left.+\frac{1}{2}\sum_{(x,x')\in\mathcal{X}^2}\tilde{a}_{xx',t}(m_t,u_t)\frac{\partial^2 }{\partial m_x \partial m_{x'}}\tilde{v}_t\right\}\\
\partial_t m_t+div\left( \tilde{f}_t(m_t,u_t^*) m_t\right)=\frac{1}{2}\sum_{x,x'}\partial^2_{xx'}\left(\tilde{a}_{xx',t}(m_t,u_t^*)m_t\right)\\
m_0\in\Delta(\mathcal{X}).
\end{array} \right.
$$ where $\tilde{f}_t$ is the drift and $\sigma_t\sigma'_t=\tilde{a}_t.$

\subsection{ Stochastic difference games}

Consider the stochastic difference equation in $\mathbb{R}:$

$$\left\{\begin{array}{c} \label{individual3}
{x}_{j,t_{k+1}^n}^n= {x}_{j,t_k^n}+\delta_n \sum_{i=1}^n \bar{\omega}^n_{ij} f_{\theta_j,t}({x}_{j,t_k^n}^n,{u}_{j,t_k^n}^n,{x}_{i,t_k^n})
\nonumber\\  \nonumber +
\sum_{i=1}^n \bar{\omega}^n_{ij} \sigma_{\theta_j,t}({x}_{j,t_k^n}^n,{u}_{j,t_k^n}^n,{x}_{i,t_k^n}^n)\left(\mathcal{B}_{j,t_{k+1}^n}^n-\mathcal{B}_{j,t_k^n}^n\right)
\\ \nonumber x_{j,0}^n=x_j,\ t_k^n=k\delta_n,\ k\geq 0,\ \delta_n>0,\lim_n\delta_n=0.
\end{array}\right.
$$ where $\bar{\omega}^n_{ij}$ is a weight representing the influence of player to player j's state.
We define the cumulative function $\tilde{F}^n$ as $\tilde{F}^n(t,w)=\sum_{j=1}^n \bar{\omega}_j^n \ind_{\{x_{j,t}^n\leq w\}}$ where
$x_{j,t}^n$ is the interpolated process from ${x}_{j,t_{k+1}^n}^n.$
For any $T<+\infty$ there exists $\tilde{c}_T>0$ such that
    $$
    \mathbb{E}\parallel \tilde{F}(t^n_k,.)-\tilde{F}^n(t^n_k,.)\parallel_1\leq \tilde{c}_T\left[ \parallel\tilde{F}_0-\tilde{F}^n_0\parallel_1+\frac{1}{\sqrt{n}}+\sqrt{\delta_n}\right]
    $$
    Moreover $\tilde{F}(t,.)$ is the solution of
    {\small \begin{eqnarray}
 \frac{\partial}{\partial t}\bar{F}_{\bar{\theta},t}(\bar{x})+\left[\int_w {f}_{\bar{\theta},t}(\bar{x},\bar{u},w)\frac{\partial}{\partial w}\bar{F}_{\bar{\theta},t}(w) dw   \right] \frac{\partial}{\partial \bar{x}}\bar{F}_{\bar{\theta},t}(\bar{x}) \\
 =\frac{1}{2}\frac{\partial}{\partial \bar{x}}\left[
\left(\int_w {\sigma}_{\bar{\theta},t}(\bar{x},\bar{u},w)\partial_{w}\bar{F}_{\bar{\theta},t}(w)dw\right)^2
\partial_{\bar{x}}\bar{F}_{\bar{\theta},t}(\bar{x})\right]
 \\ \bar{\theta}\in \Theta,\
 \bar{m}_0(.) \ \ \mbox{fixed}
 \end{eqnarray}}
 See   \cite{draftgerad2} for more details.
The  finite horizon cost function optimization leads to a coupled system of backward-forward equations:

$$\left\{\begin{array}{c}
v_{j,T}(x_j,m)=g(x_j,m)\\
-\partial_tv_{j,t}=\sup_{u_j}\left\{ r_{\theta_j,t}(x_j,u_j,m_{j,t})  +\langle\bar{f}_t(x_j,u_j,m_{j,t}), \partial_x v_{j,t}\rangle \right.  \\  \left.+\frac{1}{2}\bar{\sigma}^2_{\theta_j,t}(x_j,u_j,m_t)\partial^2_{xx} v_{j,t} \right\}\\
{d}\bar{x}_{\bar{\theta},t}=\int_w f_{\bar{\theta},t}(\bar{x}_{\bar{\theta},t},u_{\bar{\theta},t}^*,w)m_{t}(dw) {d}t
+\int_w \sigma_{\bar{\theta},t}(\bar{x}_{\bar{\theta}},u_{\bar{\theta,t}}^*,w),m_{t}(dw) {d}\mathcal{B}_t
\\  \bar{x}_0=q\\
\frac{\partial}{\partial t}{m}_{\bar{\theta},t}+\frac{\partial}{\partial {x}}\left[\bar{f}_{\bar{\theta},t}({x},{u}^*_t,{m}_t){m}_{\bar{\theta},t}   \right]
 =\frac{1}{2}\frac{\partial^2}{\partial {x}^2}\left[ \bar{\sigma}_{\bar{\theta},t}^2({x},{u}^*_t,{m}_t){m}_{\bar{\theta},t}\right]
 \\ \bar{\theta}\in \Theta,\
 m_0(.) \in \Delta(\mathcal{X})\\
 \bar{f}_t=\int_w f_{\bar{\theta},t}(\bar{x}_{\bar{\theta},t},u_{\bar{\theta},t}^*,w)m_{t}(dw)
\end{array}\right.
$$

\subsection{Risk-sensitive mean field stochastic games}
 A link between stochastic and deterministic
mean field viewpoints is provided by considering risk-sensitive stochastic approach.  The risk-sensitive approach consists to optimize the expectation $\mathbb{E}\left(\tilde{g}(R) \right)$ where $R$ is the traditional long-term payoff function. The certainty-equivalent expectation $e(R)$ is defined by $\tilde{g}(e(R))=\left(\mathbb{E}(\tilde{g}(R)) \right).$ When $\tilde{g}=e^{{y}\mu}$ is exponential
$e(R)=\tilde{g}^{-1}\left(\mathbb{E}(\tilde{g}(R))\right)=\frac{1}{\mu}\log\left( \mathbb{E}\left(e^{\mu {R}} \right) \right).
$
Consider the finite horizon payoff:
$$
R_{\mu}:=\frac{1}{\mu}*sign(\mu)\log\mathbb{E}\left( e^{\mu[g(x_{T+1})+\sum_{t'=t}^T r_{t'}(x_{t'},u_{t'},M^n_{t'})]} \right),$$
The {intuitive view of the risk-sensitive criterion} at zero is the following:
Taylor expansion at $\mu$ close to zero leads
$$ R_{\mu}=\mathbb{E}(R) +\frac{\mu}{2}var(R)+o({\mu^2})
$$ {\it This means that the risk-sensitive criterion takes into consideration not only the expectation but also the variance!}

 When $\mu\longrightarrow 0$ one gets the risk-neutral. Depending on the sign of $\mu,$ one gets the risk-seeking
 case or the risk-averse case. The analogue of BS-K becomes a multiplicative BS-K i.e a mean field version of the multiplicative Bellman-Shapley equation coupled with Kolmogorov equation. Denote by $v_{j,\mu,t}$ the optimal payoff of player $j$ with respect to $m.$
 $$\left\{\begin{array}{c}
\tilde{g}(v_{j,\mu,t}^*(x_t,m_t))=\max_{u\in \Delta(\mathcal{A}_j(x_t))}\left[ e^{{\mu}r_t(x_t,u,m_t)}\right.\\ \left.\sum_{x'}q_{x_tux',t}(m_t)\tilde{g}(v^*_{j,\mu,t+1}(x',m_{t+1}))\right]\\
{m}_{t+1}(x')=\sum_{\bar{x}\in\mathcal{X}}m_t(\bar{x})\mathcal{L}_{\bar{x}x'}(u^*_t,m_t)
\end{array}
\right.
$$ where $$u^*_t\in\arg\max_{u} e^{{\mu}r_t(x_t,u,m_t)}\sum_{x'}q_{x_tux',t}(m_t)\tilde{g}(v^*_{j,\mu,t+1}(x',m_{t+1})).$$

Considering individual state dynamics in the form of McKean-Vlasov,
$$
\left\{
\begin{array}{c}
dx_{j,t}^n=\left(\int_w f_t(x_{j,t}^n,u^n_{j,t},w) \left[\frac{1}{n}\sum_{i=1}^n\delta_{x_{i,t}^n}\right](dw)\right) dt\\ + \sqrt{\epsilon}\left(\int_w  \sigma_t(x_{j,t}^n,u^n_{j,t},w)\left[\frac{1}{n}\sum_{i=1}^n\delta_{x_{i,t}^n}\right](dw)\right) d\mathcal{B}_{j,t},\\
x_{j,0}\in \mathbb{R}^{k},\ k\geq 1\\
j\in\{1,2,\ldots,n\},\\
\end{array}
\right.
$$
and a risk-sensitive cost criterion
$R_j(\bar{u}_j,M^n;t,x_j,m)$
$$=\frac{1}{\mu}\log\mathbb{E}\left( e^{\mu[g_T(x_T)+\int_t^T r_s(x_{j,s},u_{j,s},M^n_s)\ ds]}\ | \ x_{j,t}=x_j,M^n_t=m \right),$$
We assume regular and bounded  coefficients and their derivation with the respect to the states and $\bar{u}_j:\ [0,T]\times \mathbb{R}^k\longrightarrow \mathcal{U}_j$ is piecewise continuous in $t$ and Lipschitz in $x.$
The mean field system becomes
HJBF +Fokker-Planck-Kolmogorov equation + macroscopic McKean-Vlasov  individual dynamics, i.e.,
$$\left\{\begin{array}{c}
d{x}_{j,t}=\left(\int_w f_t(x_{j,t},u_{j,t}^*,w) m_{t}(dw)\right) dt\\ + \sqrt{\epsilon}\left(\int_w  \sigma_t(x_{j,t},u_{j,t}^*,w)m_{t}(dw)\right) d\mathcal{B}_{j,t},\\
x_{j,0}=x\\
\partial_tv_{j,t}+\sup_{u_j}\left\{ \partial_{x}v_{j,t}. f_t+\frac{\epsilon}{2}\textrm{tr}(\sigma_t\sigma'_t\partial^2_{xx}v_{j,t})\right. \\
\ \ \ \ \ \ \ \ \ \ \ \left.+\frac{\epsilon\mu}{2}\parallel \sigma_t\partial_x v_{j,t}\parallel^2+r_t\right\}=0,
\\
x_j:=x; v_{j,T}(x,m)=g_T(x,m)\\
\partial_t m_t+D_x^1\left(m_t\int_w f_t(x,u^*_t,w)m_t(dw)\right)\\
=\frac{\epsilon}{2}D^2_{xx}\left(m_t\left(\int_w \sigma'_t(x,u^*,w)m_t(dw)\right)\cdot \right.
\\
\ \ \ \ \ \ \ \ \ \ \ \ \ \left. \left(\int_w \sigma_t(x,u^*,w)m_t(dw)\right)\right)
\end{array}
\right.
$$ {\it Under specific structures of drift, payoff and volatility functions, existence result can be derived using fixed point theory. Also uniqueness issue can be addressed under monotonicity conditions. However, the existence and the uniqueness conditions of the above system under general structure remain a challenging problem}.

Here $f_t(.)\in\mathbb{R}^k$ which we denote by $(f_{k',t}(.))_{1\leq k'\leq k}.$ Let
$$\underline{\sigma}_t[x,u^*_t,m_t]=\int_w \sigma_t(x,u^*_t,w)m_t(dw), $$ $\Gamma_t(.):=\underline{\sigma}_t(.)\underline{\sigma}'_t(.)$ is a square matrix with dimension $k\times k.$ The term $D^1_x(.)$ denotes
$$\sum_{k'=1}^k \frac{\partial}{\partial x_{k'}}\left(m_t\int_w f_{k',t}(x,u^*_t,w)m_t(dw)\right),$$
and the last term on $D^2_{xx}(.)$ denotes
$$
\sum_{k''=1}^k\sum_{k'=1}^k \frac{\partial^2}{\partial x_{k'}\partial x_{k''}}\left(m_t\Gamma_{k'k'',t}(.)\right).
$$
One can show that the asymptotic large
deviations results as $\mu\longrightarrow 0,$ are typically described through a  risk-neutral mean field problem. This approach  is closely related to large-deviation,  $H_{\infty}-$control, the $\min\max$ Hamiltonian of Isaacs and {\it robust mean field stochastic game}. Preliminary results can be  found in \cite{ifac11}. The model can be extended to include random switching (jump and drift process) and  delayed state measurement.
\subsection{Explicit solutions of MFSG}
There are few classes of mean field stochastic games in which explicit solutions has been found:
\begin{itemize}
\item Mean field difference games with linear states and quadratic Hamiltonian
\item Linear-quadratic mean field differential games,
\item Mean-Variance mean field differential games
\item Mean-Variance mean field difference games
\item Risk-sensitive mean field games with exponentiated long-term quadratic loss and linear dynamics.
\end{itemize}
More details can be found in \cite{meanfieldlectures}.

\subsection{ Other extensions}
$\bullet$ Extension to Poisson point processes, Levy flights, Feller processes etc.

$\bullet$ {\it Learning in large populations}
Assume that the strategy of each player is revised according to some dynamics which can be  class-dependent drift and class-dependent diffusion terms. Then, the limiting of the learning process fall down into mean field PDE. When the diffusion is zero, one get the so-called continuity equation or transport equation. We refer the reader to \cite{learninglectures,robustlearningtac,comcom2011} for recent developments on {\it combined fully distributed payoff and strategy reinforcement learning} (CODIPAS-RL).

$\bullet$ Imperfect state measurement:
Now, we assume that the state $x_{j,t}$ is not observed by player $j,$ but  $\bar{y}_{j,t}$ which is an output function of the state and noise. Under such situations, a fundamental question is:   how to estimate the state under imperfect measurement using mean field stochastic games?

 $\bullet$ Mean field stochastic games with correlated populations, different types of players including major, minor and medium players, neighborhood based partial monitoring, hierarchical structure, and dynamic conjectural variations.

$\bullet$ Mean field cooperative games; mean field network formation games; mean field Stackelberg games, mean field Bayesian games etc. Mean field Q-learning, Mean field H-learning: heterogeneous, hybrid, cost of learning, random updates, noisy strategy in large-scale systems etc.

$\bullet$ Mean field games under fractional Brownian motion, anomalous diffusion (subdiffusion, superdiffusion).

\section{Conclusions} \label{t3sec4}
In this paper we have presented recent advances in mean field stochastic games, their applications as well as their connections to related field in large-scale systems.
Below we point out some {limitations and open issues} for future works:
\begin{itemize} \item What about a system with small size (5, 7, 29, 31 players)
?
\item The curse of dimensionality problems are transformed into a
condensed form (using localized density or aggregative terms). Are we able to solve the resulting
continuum variables? What is the complexity in solving the continuum model?
\item Is there a performance loss by using mean field approach? What is the performance gap?
\item Beyond the indistinguishability per class property, what is the class of finite games for which the mean field approach  can be applied? How big is this class of games compared to the set of all games?
\end{itemize}

\bibliographystyle{plain} 
\bibliography{biblioMF1camera}

\end{document}